\documentclass[iop]{emulateapj}
\usepackage{xcolor}
\usepackage{amsmath}
\usepackage{natbib}
\usepackage{multirow}
\usepackage{bm}		
\usepackage{soul}	
\usepackage[utf8]{inputenc}
\definecolor{lila}{rgb}{0.4,0,1}
\definecolor{grau}{rgb}{0.5,0.5,0.5}
\definecolor{darkblack}{rgb}{0.2,0.2,1}
\definecolor{orange}{rgb}{1,0.5,0}

\newcommand{\MJ}{M_{\rm J}}
\newcommand{\RJ}{R_{\rm J}}

\newcommand{\RObj}{R_{\rm obj}}
\newcommand{\fig}{Figure$\:$}

\newcommand{\ls}{\rule{0pt}{10pt}}

\shorttitle{Material Properties for Massive Giant Planets and Brown Dwarfs}
\shortauthors{Becker et al.}

\begin{document}

\title{Material Properties for the Interiors of Massive Giant Planets and Brown Dwarfs}

\author{Andreas Becker$^1$, Mandy Bethkenhagen$^{1,2}$, Clemens Kellermann$^1$, Johannes Wicht$^3$ and Ronald Redmer$^1$}
\affil{$^1$Institut f\"ur Physik, Universit\"at Rostock, 18051 Rostock, Germany\\
$^2$Lawrence Livermore National Laboratory, Livermore, CA 94550, USA\\
$^3$Max-Planck-Institut f\"ur Sonnensystemforschung, 37077 G\"ottingen, Germany}

\begin{abstract}
\noindent
We present thermodynamic material and transport properties for the extreme conditions prevalent in the interiors of massive giant planets and brown dwarfs. 
They are obtained from extensive \textit{ab initio} simulations of hydrogen-helium mixtures along the isentropes of three representative objects.
In particular, we determine the heat capacities, the thermal expansion coefficient, the isothermal compressibility, and the sound velocity. Important 
transport properties such as the electrical and thermal conductivity, opacity, and shear viscosity are also calculated. Further results for associated 
quantities including magnetic and thermal diffusivity, kinematic shear viscosity, as well as the static Love number $k_2$ and the equidistance are presented. 
In comparison to Jupiter-mass planets, the behavior inside massive giant planets and brown dwarfs is stronger dominated by degenerate 
matter. We discuss the implications on possible dynamics and magnetic fields of those massive objects. The consistent data set compiled here may 
serve as starting point to obtain material and transport properties for other substellar H-He objects with masses above one Jovian mass and finally may be used 
as input for dynamo simulations.
\end{abstract}

\keywords{brown dwarfs --- conduction --- equation of state --- planets and satellites: interiors --- planets and satellites: magnetic fields --- dense matter}

\section{Introduction}\label{sec:Intro}

The number of identified exoplanets and brown dwarfs has grown substantially over the past two decades. Even though the observations rarely go beyond characterizing their mass 
and radius, they have also started to reveal additional information. In particular, the observation of global magnetic fields can offer important constraints on the interior 
dynamics. While detecting the magnetic fields of exoplanets has proven elusive so far, brown dwarfs show radio emissions clearly indicative of an internal dynamo process 
\citep{Reiners2010}. Recently, the Zeeman line broadening measurement for a brown dwarf constrained the surface field strength to about $0.5\,$T, a value consistent with the 
estimates based on the radio emissions \citep{Berdyugina2017}. Dynamo action requires an electrically conducting and convecting region but also depends on the rotation rate 
and luminosity of an object.

Numerical models for the thermal evolution, interior dynamics, or magnetic field generation are indispensable for predicting, interpreting, and understanding the observations. These 
simulations require an internal model of the studied object that also includes the transport properties. 
Early approaches to determine the transport properties of degenerate matter were, for example, based on the Kubo theory \citep{Hubbard1966,Hubbard1969}. 
\cite{Flowers1976} used a variational approach for the solution of the Boltzmann equation and considered all relevant scattering mechanisms. \cite{Stevenson1977a} provided 
approximate formulae based on the Ziman theory using the static ion-ion structure factor for a hard-sphere system which can be applied for a wide range of densities and 
temperatures. \cite{Nandkumar1984} relied on the relaxation time approximation and included structure factor effects within the simple one-component plasma model. This 
approach has been adapted recently by \cite{Harutyunyan2016} to determine the electrical conductivity in the warm crusts of neutron stars. Conductivity models that are valid 
for the wide ranges of density and temperature in astrophysical and other applications, such as inertial confinement fusion, were proposed by, e.g., \cite{Lee1984} and 
\cite{Ichimaru1985}.

Our work follows a different path to describe the extreme matter in the interior of massive objects governed by strongly correlated ions immersed in a degenerate electron gas. We
apply a combination of density functional theory for the electron system and classical molecular simulations for the ions (DFT-MD method) to derive the material properties of
H-He mixtures. This approach had been previously applied to Jupiter by \citet{French2012}, whose results were subsequently used for simulations of the planet's interior dynamo 
processes reproducing the Jovian large scale magnetic field \citep{Gastine2014,Jones2014,Duarte2018}. 

The work presented here extends the Jupiter study of \cite{French2012}. We select three objects within a mass range of 
$10-50~\MJ$ from \cite{Becker2014}: the massive exoplanet KOI-889b, and the two brown dwarfs Corot-3b and Gliese-229b. The latter is the most massive object in this set. 
\cite{Becker2014} predict a pressure of 22000~TPa, a temperature of 1.2$\cdot$10$^6$~K, and a density of 450~g/cm$^3$ at the center of Gliese-229b based on \textit{ab initio} 
equations of state (EOS). This exceeds the thermodynamic conditions within Jupiter several orders of magnitude. For instance, the core-mantle boundary in Jupiter is predicted at 4~TPa, 
20000~K, and 4.3~g/cm$^3$ \citep{Guillot1999,Nettelmann2012}. 

The thermodynamic conditions typical for the interior of Gliese~229b are already accessible in the laboratory using the world's most powerful laser at the National Ignition Facility 
(NIF) covering the conditions from substellar objects (giant planets, brown dwarfs) to low-mass stars~\citep{Lindl2004, Moses2011}. It has been demonstrated by~\cite{Hurricane2014}
that a deuterium-tritium capsule can be dynamically compressed up to about 400~g/cm$^3$. Hence, the data presented here may also serve as input for the hydrodynamic simulations
accompanying these experiments.

Our paper is organized as follows. We recapitulate the calculation of interior models for massive giant planets and brown dwarfs according to \cite{Becker2014} in 
section~\ref{sec:IntBrownie}. The determination of the thermodynamic material properties via \textit{ab initio} simulations is outlined in section~\ref{sec:Tdyn_Props}. 
The Love number $k_2$ and the equidistance are discussed in section~\ref{sec:k2}. The calculations of the transport properties and corresponding results are
described in section~\ref{sec:Lik}. Finally, we discuss the implications of our obtained material properties on planetary and stellar dynamos in section \ref{sec:Implications} 
and conclude in section \ref{sec:Conclusion}.

\section{Interior Structure Models and Temperature Profiles}\label{sec:IntBrownie}

The interior structure models of the massive giant planet KOI-889b and the brown dwarfs Corot-3b and Gliese-229b are adopted from \cite{Becker2014}, whose assumptions we 
recall briefly.

The objects are described as spherical bodies, each composed of a single isentropic layer. The temperature profile can be obtained by integrating numerically the differential 
equation

\begin{eqnarray}
\left(\frac{\partial T}{\partial \varrho}\right)_{s}=\frac{T}{\varrho^2}\frac{\left(\frac{\partial P}{\partial T}\right)_{\varrho}}{\left(\frac{\partial u}{\partial T}\right)_{\varrho}}  
\label{IsenDGL}
\end{eqnarray}

at constant specific entropy $s$. Most crucial inputs are the thermal and caloric EOS, $P(\varrho, T)$ and $u(\varrho, T)$, which describe pressure $P$ and specific 
internal energy $u$ in terms of density $\varrho$ and temperature $T$. The material inside each object is approximated as linear mixture of hydrogen, helium and heavier elements (with their 
respective mass fractions $X$, $Y$ and $Z$). Here, we use H-REOS.3 for hydrogen and He-REOS.3 for helium,  which have been derived from \textit{ab initio} simulations employing 
VASP~\citep{Kresse1993,Kresse1994,Kresse1996,Hafner2008}, see~\cite{Becker2014} for details. The heavy elements are approximated with a fourfold mass-scaled version of He-REOS.3 to 
match the mass of water. In the modeling procedure, $Y$ is fixed to the solar value of 0.27, while $Z$ is varied until the resulting model matches the observational constraints for total 
mass and radius. $X$ is given by $X=1-Y-Z$. The final compositional triplets $X/Y/Z$ can be found in Table~\ref{tab:objects} together with the observational constraints and the boundary
conditions of each isentrope. The boundary of each object is chosen at pressure $P_{atm}$ and temperature $T_{atm}$, where the atmosphere becomes convective and therefore marks the onset 
of the isentrope. \cite{Becker2014} employed the radiative-convective model of \cite{Marley1996} and \cite{Fortney2008b} to describe the objects' atmospheres.

The interior structure models by \cite{Becker2014} result in a mass $M$ of $9.98~\MJ$ and a radius $R$ of $1.028~\RJ$ for KOI-889b, $21.66~\MJ$ and $0.973~\RJ$ for Corot-3b , 
and $46.23~\MJ$ and $0.8646~\RJ$ for Gliese-229b. All these values lie within the error bars of the corresponding observational constraints as listed in Table~\ref{tab:objects}. 

\begin{deluxetable}{cccc}[bht]
\tablecolumns{4}
\tablewidth{0pc}
\tablecaption{Observational constraints and parameters}
\tablehead{ \colhead{~}
  & \colhead{KOI-889b\tablenotemark{a,d}}
  & \colhead{Corot-3b\tablenotemark{b,d}}
  & \colhead{Gliese-229b\tablenotemark{c,d}} }
\startdata
\ls Mass [$\MJ$]         &$9.98 \pm 0.5$   &$21.66 \pm 1$    &$46.2_{-14.8}^{+11.8}$  \\
\ls Radius [$\RJ$]       &$1.03 \pm 0.06$  &$1.01 \pm 0.07$  &$0.87_{-0.07}^{+0.11}$  \\
\ls Fe/H                 &-0.07$\pm$0.15   &-0.02$\pm$0.06   &-0.2$\pm$0.4            \\
\ls P$_{\rm atm}$ [bar]  &58               &74               &52                      \\
\ls T$_{\rm atm}$ [K]    &1000             &1500             &1800                    \\
\ls X/Y/Z	     	 &0.69/0.27/0.04   &0.71/0.27/0.02   &0.71/0.27/0.02	      \\
\enddata
\tablenotetext{a}{Mass, radius, and Fe/H: \cite{Hebrard2013}}
\tablenotetext{b}{Mass, radius, and Fe/H: \cite{Deleuil2008}}
\tablenotetext{c}{Mass and radius are derived from the fitting formulae given in \cite{Marley1996}, Fe/H is taken from \cite{Schiavon1997}.}
\tablenotetext{d}{P$_{\rm atm}$, T$_{\rm atm}$, and X/Y/Z: \cite{Becker2014}, atmosphere model by \cite{Marley1996} and \cite{Fortney2008b}}
\tablenotetext{ }{ }
\label{tab:objects}
\end{deluxetable}

\begin{figure}[hbt]
  \includegraphics[width=1.0\columnwidth]{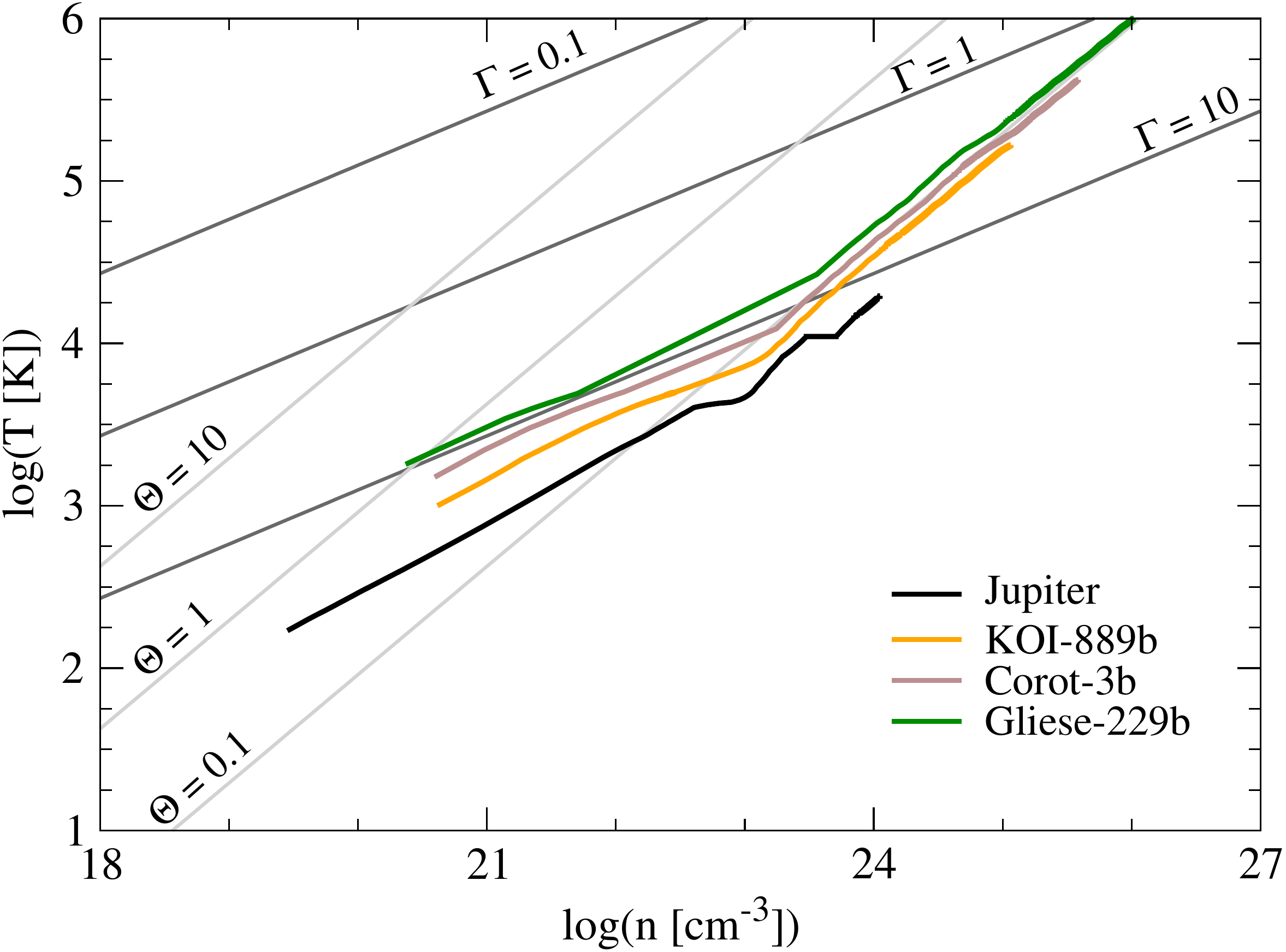}
  \caption{Temperature-particle density plane with isentropes of the three massive objects under consideration here (see \cite{Becker2014}) in comparison to Jupiter (see 
  \cite{Nettelmann2012,French2012}). Also shown are lines for the ionic coupling parameter $\Gamma=0.1, 1, 10$ and the electron degeneracy parameter $\Theta=0.1, 1, 10$ which 
  illustrate the thermodynamic conditions along the isentropes and the parameter region for our present calculations.}
  \label{fig:PlasmaParam}
\end{figure}

The isentropes of the three objects considered here are illustrated in Figure~\ref{fig:PlasmaParam}, while the thermodynamic states for selected points along the isentropes 
can be found in Table~\ref{tab:Mat}. For comparison we also include the three-layer Jupiter model J11-8a with an outer enevelope composition $X_1/Y_1/Z_1$ of 0.724/0.238/0.038 
and an inner envelope composition $X_2/Y_2/Z_2$ of 0.561/0.311/0.128 \citep{Nettelmann2012, French2012} in Figure~\ref{fig:PlasmaParam}.
Additionally, we show the coupling parameter $\Gamma=e^2/(4\pi\varepsilon_0 d k_BT)$ and the degeneracy parameter $\Theta=k_BT/E_F$ associated with the interior temperatures for the case 
of hydrogen-helium mixtures of equal electron and ion number density. $d$ is the mean distance between the ions, $k_B$ represents Boltzmann's constant, and $E_F$ the electronic Fermi 
energy. 
It can be seen that even in the case of the object with the smallest mass, KOI-889b, the 
particle density and temperature in the center are about one order of magnitude higher than in Jupiter. All four isentropes are subject to strong electron degeneracy
($\Theta\sim 0.1$) in the inner regions. The pronounced change in slope in all four curves is due to dissociation and ionization of hydrogen. The density jump at about 
$5\times 10^{23}$~cm$^{-3}$ on the Jupiter isentrope marks the boundary between the upper and lower mantle, see~\cite{Nettelmann2012}.

\section{Thermodynamic Material Properties}\label{sec:Tdyn_Props}

The entire set of thermodynamic material properties is directly derived from the linearly mixed wide-range EOS for hydrogen and helium~\citep{Becker2014} described in the previous 
section. Additional data points were generated via cubic spline interpolation of the EOS tables to obtain 
a sufficiently dense EOS grid. The derivatives of $P(\varrho,T)$ and $u(\varrho,T)$ and thus the material properties were then calculated analytically from the spline polynomials. 
Consequently, the errors of our results are connected to the errors of the EOS tables and are estimated to be $\leq 5\%$, see~\cite{Becker2014}. 

In this section we present results for the specific heat capacities at constant volume $c_V$ and constant pressure
$c_P$, as well as the isothermal compressibility $\kappa_T$, the thermal expansion coefficient $\alpha$, and the sound
velocity $c_s$. These quantities are defined as follows:
\begin{eqnarray}
 \alpha & = & -\frac{1}{\varrho}\left(\frac{\partial \varrho}{\partial T}\right)_P =\frac{1}{\varrho} \left(\frac{\partial \varrho}{\partial P}\right)_T\left(\frac{\partial P}{\partial T}\right)_\varrho,\\
 \kappa_T &=&  \frac{1}{\varrho}\left(\frac{\partial \varrho}{\partial P}\right)_T ,\\
  c_V & = &\left(\frac{\partial u}{\partial T}\right)_{\varrho} ,\\
 c_P & = & c_V + \frac{T\alpha^2}{\varrho\kappa_T},\label{eq:CpCv} \label{eq:cpcv} \\
 c_s &=& \sqrt{\left(\frac{\partial P}{\partial\varrho}\right)_s}= \sqrt{\frac{c_P}{\varrho c_V\kappa_T}} .\label{eq:TdynMatConst}
\end{eqnarray}
The respective values along the isentropes of the three considered objects are summarized in Table~\ref{tab:Mat}.

\begin{figure}[htb]
  \includegraphics[width=1.0\columnwidth]{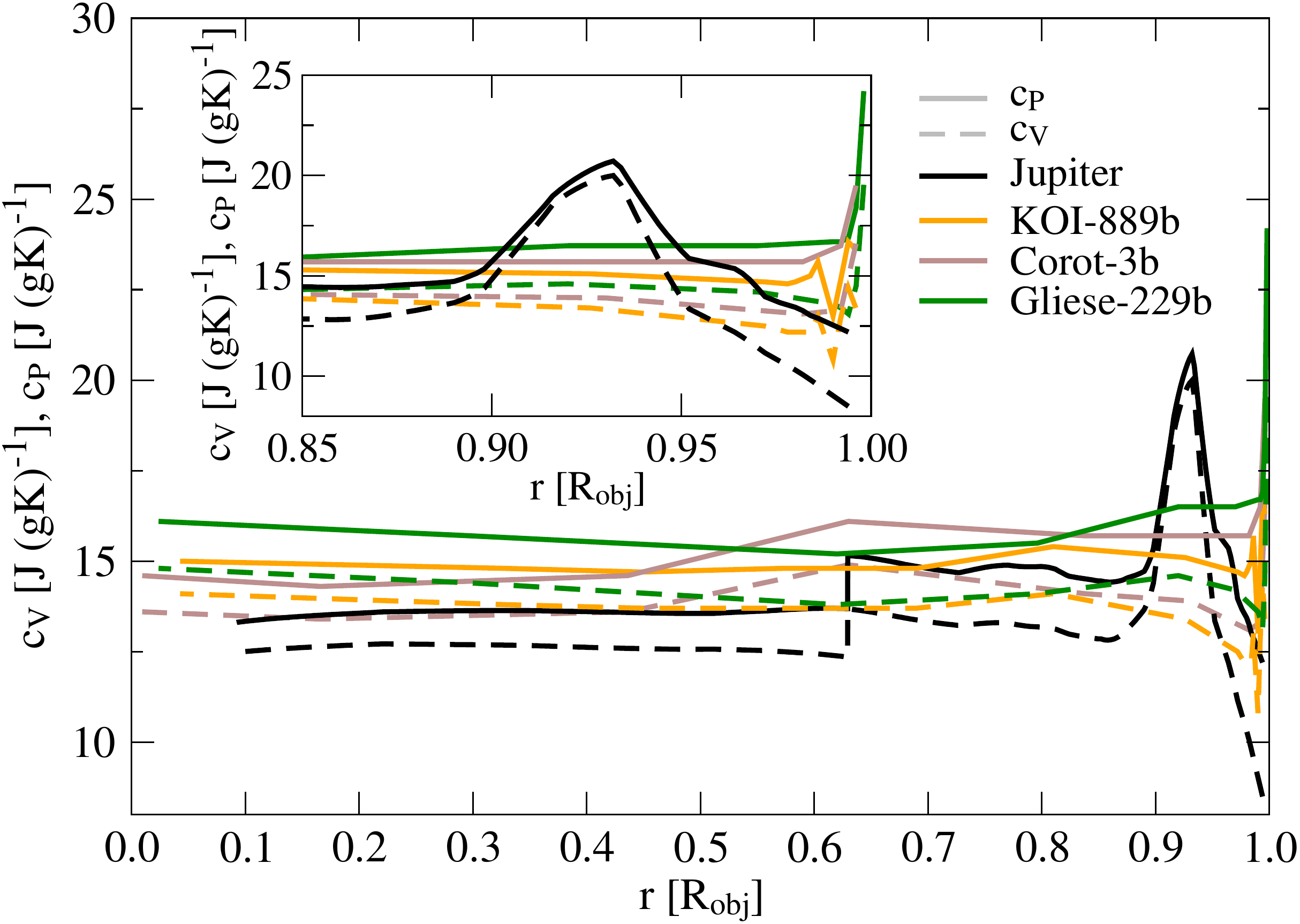}
  \caption{Specific heat capacities at constant volume $c_V$ (dashed curves) and constant pressure $c_P$ (solid curves) along the isentropes of Jupiter 
  (black)~\citep{French2012}, KOI-889b (orange), Corot-3b (brown) and Gliese-229b (green).}
  \label{fig:CpCv}
\end{figure}

\begin{figure}[htb]
  \includegraphics[width=1.0\columnwidth]{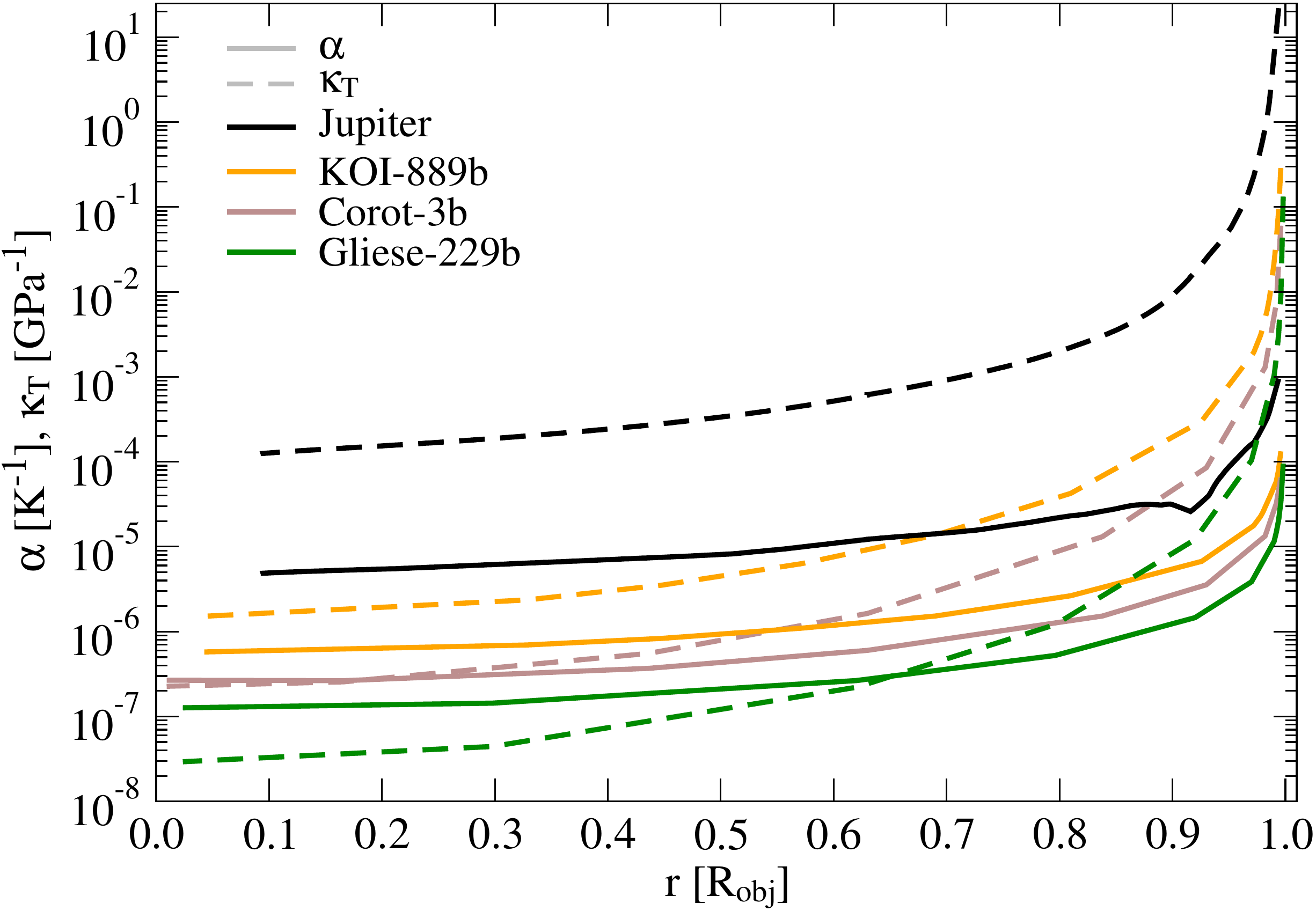}
  \caption{Isobaric expansion coefficient $\alpha$ (solid curves) and the isothermal compressibility $\kappa_T$ (dashed curves) along the considered
  isentropes with the same color code used in Figure~\ref{fig:CpCv}.}
  \label{fig:kappaT}
\end{figure}

\begin{figure}[htb]
  \includegraphics[width=1.0\columnwidth]{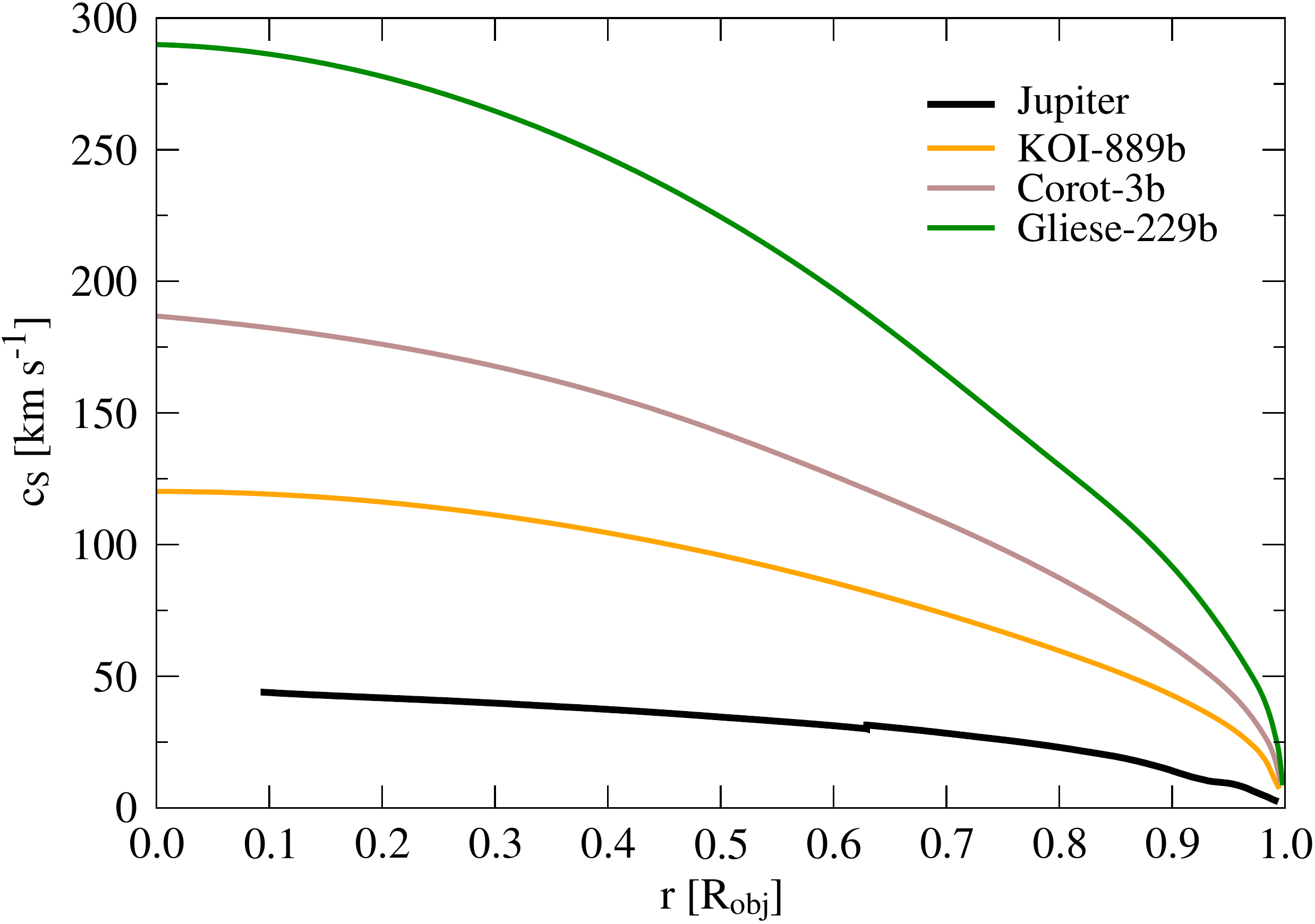}
  \caption{Sound velocities along the isentropes of Jupiter (black)~\citep{French2012}, KOI-889b (orange), Corot-3b (brown), and Gliese-229b (green).}
  \label{fig:Cs}
\end{figure}

\begin{deluxetable*}{ccccccccccc}[htb]
\tablecolumns{11}
\tablewidth{0pc}
\tablecaption{Thermodynamic material properties of the massive giant planet KOI-889b as well as the brown dwarfs Corot-3b and Gliese-229b}
\tablehead{
\colhead{object} & \colhead{r} & \colhead{m} & \colhead{P} & \colhead{T} & \colhead{$\varrho$} & \colhead{$\alpha$} 
& \colhead{$\kappa_T$} & \colhead{c$_V$} & \colhead{c$_P$} & \colhead{c$_s$}\\
\colhead{} & \colhead{[R$_{\mathrm{obj}}$]} & \colhead{[M$_{\mathrm{obj}}$]} & \colhead{[GPa]} & \colhead{[K]} & \colhead{[g cm$^{-3}$]} & \colhead{[K$^{-1}$]} 
& \colhead{[GPa$^{-1}$]} & \colhead{[J (gK)$^{-1}$]} & \colhead{[J (gK)$^{-1}$]} & \colhead{[km s$^{-1}$]}}
\startdata
KOI-889b &	0.996		& 0.99996		& 2.53			& 4800		& 0.100		& 1.33$\times10^{-4}$	& 0.288			& 13.4	& 16.3	& 6.48\\
\vdots & \vdots &\vdots &\vdots &\vdots &\vdots &\vdots &\vdots &\vdots &\vdots & \vdots\\
KOI-889b &	0.043		& 3.15$\times10^{-4}$	& 3.974$\times10^{5}$	& 166000	& 46.5		& 5.75$\times10^{-7}$	& 1.51$\times10^{-6}$	& 14.1	& 15.0	& 120.0\\
Corot-3b &	0.996			& 0.9558		& 11.27			& 7700		& 0.208		& 6.53$\times10^{-5}$	& 0.06			& 16.6	& 19.5	& 9.60\\
\vdots & \vdots &\vdots &\vdots &\vdots &\vdots &\vdots &\vdots &\vdots &\vdots & \vdots\\
Corot-3b &	9.29$\times10^{-3}$	& 3.08$\times10^{-5}$	& 2.7600$\times10^{6}$	& 420000	& 136.6		& 2.67$\times10^{-7}$	& 2.27$\times10^{-7}$	& 13.6	& 14.6	& 186.5\\
Gliese-229b &	0.998		& 1.0000		& 5.76			& 8000			& 0.126		& 9.50$\times10^{-5}$	& 0.132			& 19.5	& 24.2	& 8.59\\
\vdots & \vdots &\vdots &\vdots &\vdots &\vdots &\vdots &\vdots &\vdots &\vdots & \vdots\\
Gliese-229b &	0.024		& 6.84$\times10^{-5}$	& 2.214$\times10^{7}$	& 1.18$\times10^{6}$	& 447		& 1.26$\times10^{-7}$	& 2.93$\times10^{-8}$	& 14.8	& 16.3	& 289\\
\enddata
\tablenotetext{}{(This table is available in its entirety in machine-readable form.)}
\label{tab:Mat}
\end{deluxetable*}

The results for the specific heat capacities $c_V$ (dashed curves) and $c_P$ (solid curves) for Gliese-229b (green), Corot-3b (brown), and KOI-889b (orange) are shown in 
Figure~\ref{fig:CpCv}. The heat capacities for Jupiter (black), computed earlier by~\cite{French2012}, are shown for comparison. Note, no data are provided for the innermost 
region within 10\% of Jupiter's radius, which is assumed to be an isothermal rocky core. For Jupiter, both heat capacities are characterized by a pronounced peak 
at $90\%$ of its radius, which coincides with the dissociation of the hydrogen molecules and the associated latent heat. The step in $c_V$ and $c_P$ at $\sim 62\%$ of the radius 
is due to the transition from the outer to the inner envelope layer. 
The three more massive objects lack such a discontinuity, since they are modeled with one layer only. Moreover, we observe no pronounced local maxima as found in 
the case of Jupiter. This is due to the much hotter boundary temperatures for the isentropes of the massive giant planet KOI-889b and the brown dwarfs.
While the Jupiter model starts at 1000 K where hydrogen is still in the molecular state, 
\cite{Becker2014} determined boundary temperatures between 4800~K and 8000~K for the massive objects where most of the hydrogen molecules are already dissociated.
The corresponding heat capacity curves thus only capture the falling flanks of the peaks caused by the latent heat of hydrogen dissociation. The respective radial gradients in 
Figure~\ref{fig:CpCv} seem steeper than in Jupiter because we use relative rather than absolute radii.
At deep interior conditions, all profiles remain roughly constant. This reflects the simpler properties of the fully ionized and increasingly degenerate matter described 
by~\cite{Hubbard1966}. However, the particles are still significantly correlated under these conditions, e.g. 16\% of the heat capacity $c_V$ can be attributed solely interaction 
effects in Gliese-229b's interior.

The close approach of $c_V$ and $c_P$ at the maximum of the Jovian curves can be explained using Equation~(\ref{eq:cpcv}), where the isobaric expansion coefficient $\alpha$
contributes quadratically.
This quantity is compared to the isothermal compressibility $\kappa_T$ in Figure~\ref{fig:kappaT}. All curves decrease entirely monotonically with density,
with the exception of a minimum in the $\alpha$ profile of Jupiter due to the dissociation of hydrogen. This minimum in turn causes $c_V$ and $c_P$ to be
particularly similar at this point (see Equation~(\ref{eq:cpcv})).
Overall, we find $\kappa_T$ and $\alpha$ to be smaller for more massive objects because of the higher degeneracy of the matter in their interiors.

The sound velocity $c_s$ can be derived from the properties discussed above using Equation~(\ref{eq:TdynMatConst}). The resulting profiles are shown in Figure~\ref{fig:Cs}.
For all considered objects, the sound velocity increases with $\varrho$, since $\kappa_T$ decreases with density steeper than 1/$\varrho$ while $c_V/c_P$ remains roughly 
constant.

\section{Static Love Number and Equidistance}\label{sec:k2}

While the material properties of the matter inside planetary objects are usually not directly accessible, we can at least narrow down
the mass distribution inside an object by measuring the Love number $k_2$ and equidistance $\nu_0$. Therefore, we provide values for all objects
under consideration and compare to experimental values in Table~\ref{tab:love-equi}. The numerical procedure is adapted from \cite{Kellermann2018} and
is briefly summarized in the following.

If a planet is in the vicinity of another mass, e.g., the parent star or a moon, its gravitational field will be perturbed by the interactions.
This external potential can be expanded using Legendre polynomials $P_n$:
\begin{eqnarray}
  W(s) = \sum_{n=2}^\infty W_n = \frac{GM}{a} \sum_{n=2}^\infty \left(\frac{s}{a}\right)^n P_n(\cos\theta) \quad,
\end{eqnarray}
with the perturbing mass $M$, its distance to the planet $a$, the radial coordinate within the planet $s$, and
the angle $\theta$ between $s$ and $a$. Due to this tidal mass shift, the planet's potential responds with the induced potential
\begin{eqnarray}
  V_n^{\rm ind}(s) &=& K_n(s)W_n(s) \quad,\\
  V_n^{\rm ind}(\RObj) &=& k_n W_n(\RObj) \quad.
\end{eqnarray}
$K_n$ is the Love function of degree $n$ and its value at the surface is the Love number $k_n$.
To obtain $k_n$ we follow the formulation of \cite{Zharkov1978}. Similar to the gravitational moments $J_{2n}$,
the Love numbers depend on the density profile of the planet. They can be calculated via
\begin{eqnarray}
  k_n = \frac{T_n(\RObj)}{\RObj ~ g_0}-1 \quad.
\end{eqnarray}
Here $g_0$ is the surface gravity of the unperturbed planet and the function $T_n$ fulfills the differential equation
\begin{eqnarray}
  T_n''(s) + \frac{2}{s} T_n'(s) + \left[ \frac{4\pi G \varrho'(s)}{V'(s)} - \frac{n(n+1)}{s^2} \right] T_n(s) = 0  \quad,\nonumber\\
\end{eqnarray}
with $\varrho$ and $V$ the density profile and potential of the unperturbed planet, respectively. Primed quantities denote derivatives with respect to 
coordinate $s$.

In the case of Jupiter's three-layer model we have to account for internal density jumps. Therefore, the inner boundary conditions
\begin{eqnarray}
  T_n(b^+) &=& T_n(b^-)  \quad,\\
  T_n'(b^+) &=& T_n'(b^-) + \frac{4\pi G}{V'(b)}\left[\varrho(b^-)-\varrho(b^+)\right]T_n(b)
\end{eqnarray}
have to be fulfilled. In these equations $b^-$ and $b^+$ denote the inner and outer side of the density jump, respectively.

The possible results for $k_2$ lie between 0 and 1.5, the latter value being the limit for a sphere with homogeneous density. In general, a concentration
of mass towards the center results in a smaller $k_2$ value.

Our calculated values for Jupiter, KOI-889b, and the two brown dwarfs are listed in Table~\ref{tab:love-equi}. The table also contains the observational value for 
Jupiter \mbox{$k_{2,{\rm J}}=0.49$} that is based on gravity measurements \citep{Ragozzine2009} and can be reproduced by the three-layer model to within 10~\%. For the three massive 
objects the mass increases strongly but the radius decreases only slightly compared to Jupiter.
While the Love number of the giant planet KOI-889b is still comparable to that of Jupiter, the significant rise in gravitational pull towards the center 
inside the brown dwarfs leads to a reduction of $k_2$ down to 0.349 for Gliese-229b.

The equidistance $\nu_0$ (characteristic frequency) is the inverse of the time an acoustic wave would need to travel through an object. It is sensitive to the internal structure
(see \cite{Gudkova1999b})
and is related to the sound velocity via 
\begin{eqnarray}
  \nu_0 &=& \left[ 2\int_0^{R_{\mathrm{Obj}}} \frac{{\rm d}r}{c_s(r)} \right]^{-1} \quad. \label{eq:equidist}
\end{eqnarray}

The equidistance thus increases with the object mass like the sound velocity (see Table~\ref{tab:love-equi}).
For Jupiter our calculated value of $156~\mu$Hz agrees well with the experimental one of $155.3\pm 2.2~\mu$Hz \citep{Gaulme2011}. Furthermore, \cite{LeBihan2013} calculated the equidistance 
for massive giant planets. For Corot-3b they find an equidistance of 653.3~$\mu$Hz which is considerably lower than 723~$\mu$Hz as derived from our model. 
However, they used a different equation of state for the hydrogen-helium system with a slightly lower helium mass fraction ($Y=0.25$) 
and any metallicity is neglected ($Z=0$). In particular, they used the upper limit for the radius of Corot-3b (1.01~$\RJ$) while our results are based on a model
predicting 0.973~$\RJ$. Thus, we obtain steeper gradients in the density profile, leading to higher sound velocities and finally to a larger value of the equidistance.

\begin{deluxetable}{ccc}[htb]
\tablecolumns{3}
\tablewidth{0pc}
\tablecaption{Love number $k_2$ and equidistance $\nu_0$}
\tablehead{ \colhead{Object} &\colhead{$k_2$} & \colhead{$\nu_0$ [$\mu$Hz]}}
\startdata
Jupiter (exp.)\tablenotemark{a,b}	& 0.49	& 155.3 $\pm$ 2.2\\
Jupiter (3L)\tablenotemark{c} 	  	& 0.538 & 156  \\
KOI-889b (1L)\tablenotemark{  }    	& 0.447 & 464  \\
Corot-3b (1L)\tablenotemark{  }    	& 0.387 & 723  \\
Gliese-229b (1L)\tablenotemark{  } 	& 0.349 & 1235 \\
\enddata
\tablenotetext{a}{Love number: \cite{Ragozzine2009}}
\tablenotetext{b}{Equidistance: \cite{Gaulme2011}}
\tablenotetext{c}{Underlying three-layer (3L) model: \cite{Nettelmann2012}}
\tablenotetext{}{}
\label{tab:love-equi}
\end{deluxetable}

\section{Transport Properties}\label{sec:Lik}

In contrast to the thermodynamic properties, the transport properties such as the shear viscosity $\eta$ and the electrical and thermal conductivities $\sigma$ and $\lambda$ are 
obtained from simulating a real H-He mixture. The DFT-MD simulations are carried out with VASP, assuming a solar-like mean helium content of $\overline{Y}=0.275$. The calculations
for the viscosity are performed with 116 hydrogen and 11 helium ions, while the static simulations for $\sigma$ and $\lambda$ require 232 hydrogen and 22 helium ions. 
The choice between the full Coulomb potential and PAW pseudopotentials~\citep{Blochl1994} and in turn the cutoff energies was made according to \citet{Becker2014}. Subsequently, the
remaining transport properties such as the Rosseland mean opacity $\kappa_R$ and the Lorenz number $L$ are derived from the conductivities. In general, all transport properties possess an
ionic as well as an electronic contribution arising from the Born-Oppenheimer approximation allowing to decouple the motion of both species. In appropriate cases one of these two 
contributions is neglected, as discussed in the following sections for the individual transport properties.

The densities and temperatures are selected along the isentropes of Gliese-229b, Corot-3b and KOI-889b, see Table~\ref{tab:Mat}. Heavier elements ($Z$) are not explicitly 
represented in those simulations, due to their small abundance. However, the helium content was increased slightly from 0.27 to 0.275 to match the 
pressures of the isentropes discussed in section~\ref{sec:IntBrownie}, which included a mass-scaled helium EOS to represent heavy elements. Indeed, the resulting pressures of 
the real mixture simulations reproduce the pressures along the isentropes with a maximum deviation of $2.5\%$ for Gliese-229b and Corot-3b (both $Z=2\%$), and $5\%$ for KOI-889b 
due to the larger amount of heavier elements ($Z=4\%$). The entire sets of transport properties along the three isentropes can be found in Table~\ref{tab:Trans}.

\subsection{Viscosities}\label{subsec:Visco_Theo}

Since the shear viscosity $\eta$ is dominated by the motion of the ions in the system, we neglect the electronic contributions, following~\cite{Bertolini2007}.
The shear viscosity can be evaluated within the framework of linear response theory (LRT) (\cite{Kubo1957, Allen1989, Alfe1998}) using autocorrelation functions (ACFs) 
for the non-diagonal elements of the stress tensor $p_{ij}$:
\begin{equation}
 \eta=\frac{V}{3k_BT}\int\limits_0^\infty\mathrm{d}t \sum_{ij=\left\{xy,yz,zx \right\}}\langle p_{ij}(0)p_{ij}(t) \rangle \quad.
 \label{eq:visco}
\end{equation}
Here, $V$ is the volume of the simulation box, $T$ the temperature and $k_B$ is Boltzmann's constant.
Following~\cite{Alfe1998}, two additional ACFs are defined via rotation invariance by a linear combination of the diagonal elements of the stress tensor:
$1/2(\langle p_{xx}(0)p_{xx}(t) \rangle - \langle p_{yy}(0)p_{yy}(t) \rangle)$ und $1/2(\langle p_{yy}(0)p_{yy}(t) \rangle - \langle p_{zz}(0)p_{zz}(t) \rangle)$.
Below 50000~K we used all five independent ACFs to calculate $\eta$ for a better statistics, while at 50000~K and above only the three ACFs given by Equation~(\ref{eq:visco}) 
were accessible.

In the respective DFT-MD calculations we sampled the Brillouin zone with a $2\times2\times2$
Monkhorst-Pack $\mathbf{k}$-point grid. Using only the Baldereschi mean value point \citep{Baldereschi1973} is not sufficient since the ACFs do not fluctuate around zero
in this case, leading to wrong results. For sufficient statistics of the ACFs and a maximum error of 15\% for $\eta$, but typically below 10\%, we simulated at least 120000
time steps, leading to total simulated times between 20 and 80~ps for each point.

\begin{figure}[htb]
  \includegraphics[width=1.0\columnwidth]{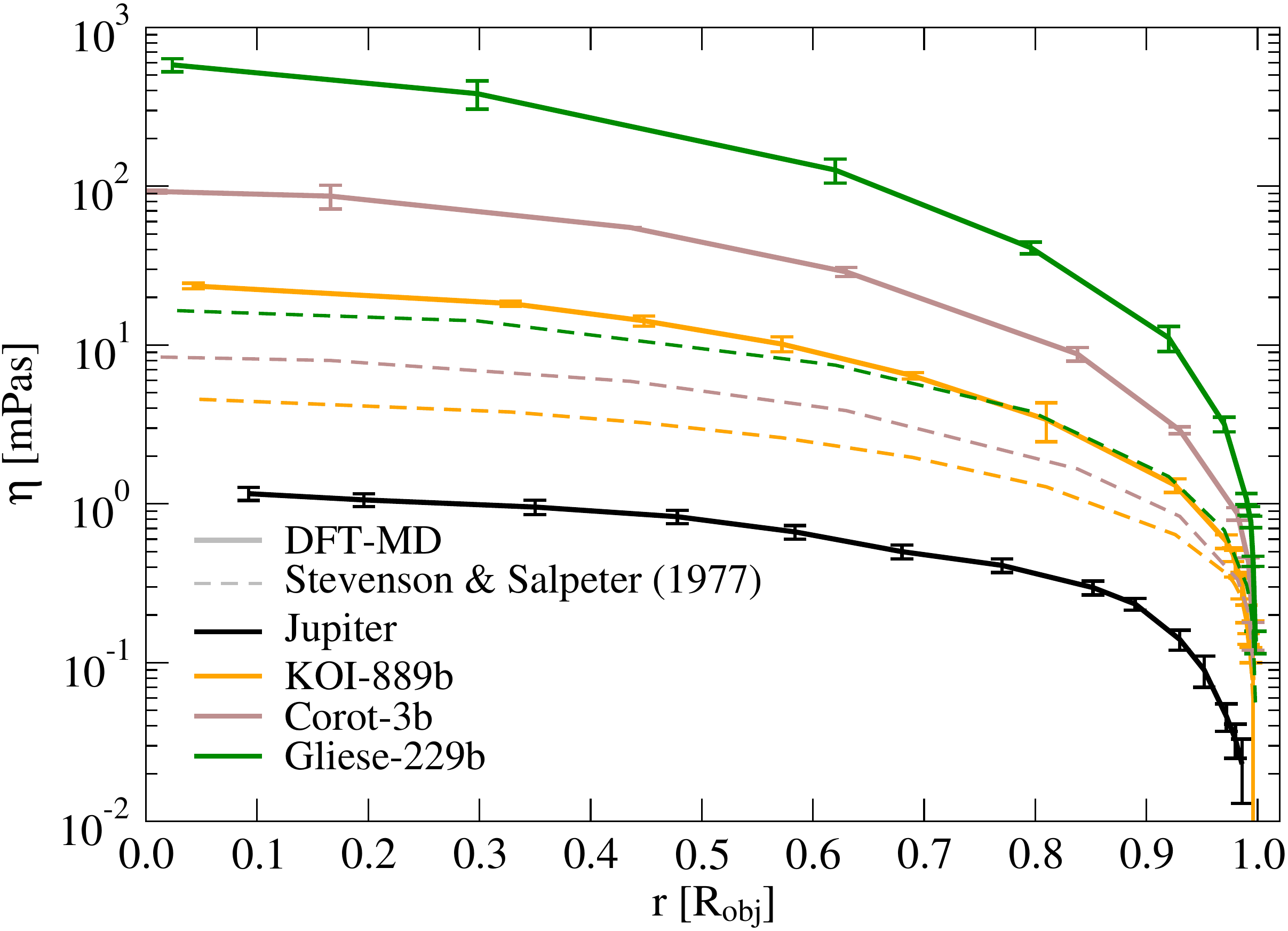}
  \caption{Dynamic shear viscosities $\eta$ from DFT-MD calculations (solid curves) for Jupiter (black)~\citep{French2012}, KOI-889b (orange), Corot-3b (brown), and 
  Gliese-229b (green) in comparison to results of~\cite{Stevenson1977a} (dashed curves).}
  \label{fig:eta1}
\end{figure}

\begin{figure}[htb]
  \includegraphics[width=1.0\columnwidth]{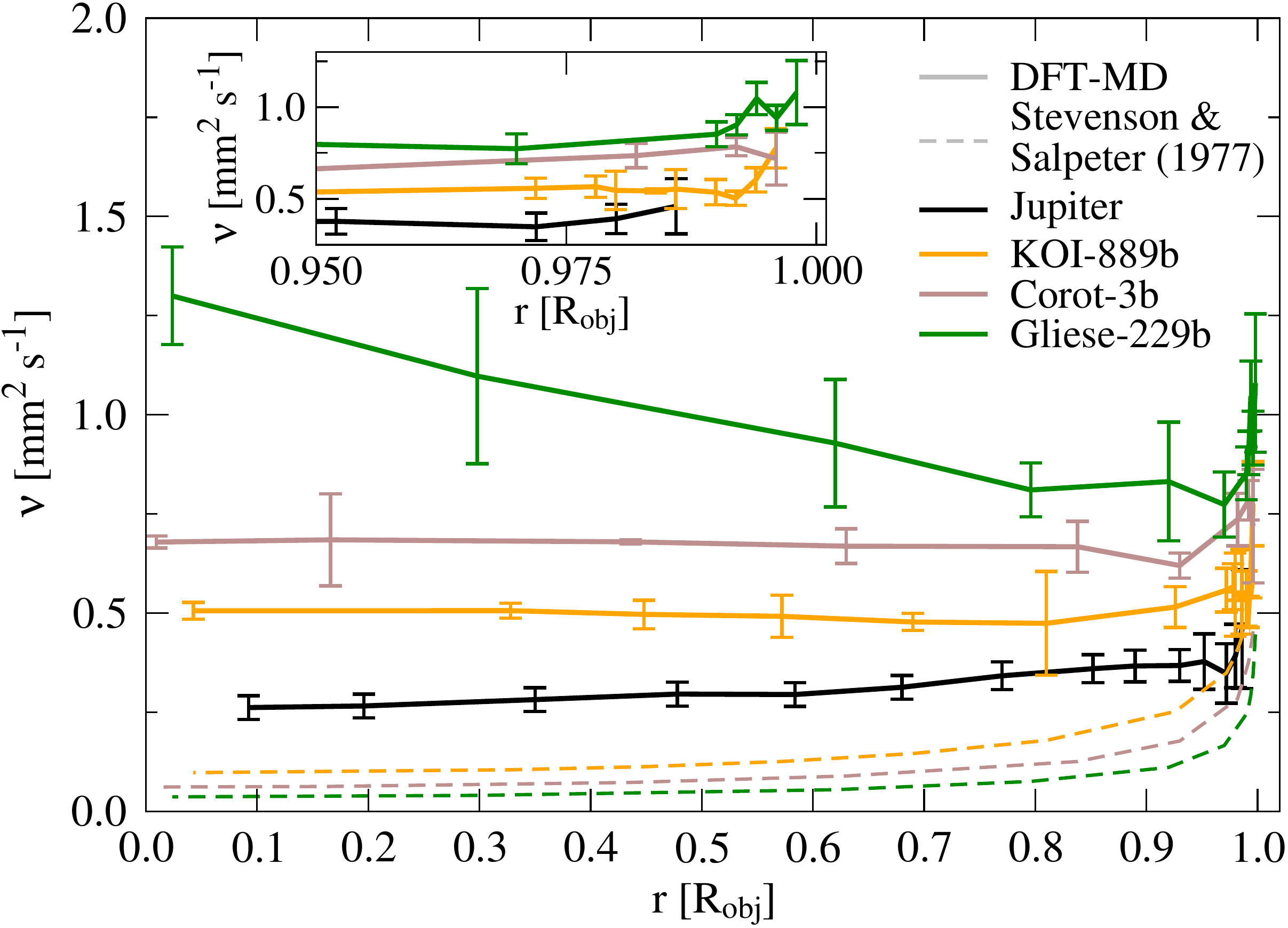}
  \caption{Kinematic shear viscosities $\nu=\eta/\varrho$ displayed in the same color code as in Figure~\ref{fig:eta1}. The results of~\cite{Stevenson1977a}
  (dashed curves) are shown for comparison.}
  \label{fig:eta2}
\end{figure}

The results for the dynamic shear viscosity are shown in Figure~\ref{fig:eta1}. The solid curves represent our DFT-MD results for KOI-889b (orange), Corot-3b (brown),
and Gliese-229b (green) along with earlier DFT-MD calculations for Jupiter (black) by \cite{French2012}. Additionally, the dashed curves illustrate predictions
of~\cite{Stevenson1977a} for comparison.  The viscosity is increasing monotonically from the surface to the center for all objects. The more compact the giant planet/brown 
dwarf is, the higher is the density inside the object and the higher is the shear viscosity. The dynamic shear viscosities provided by~\cite{Stevenson1977a} are based on 
the behavior of fluid alkali metals (see also \cite{Nandkumar1984}) and underestimate the viscosities considerably compared to the DFT-MD results. 
This behavior is also observed for the kinematic shear viscosity $\nu=\eta/\varrho$ illustrated in Figure~\ref{fig:eta2}. 
The results of \cite{Stevenson1977a} lie entirely below the DFT-MD curve of Jupiter and show a reversed trend compared to our calculations. 
We find the kinematic shear viscosities of our considered objects to increase with higher mass. Moreover, the values for KOI-889b and 
Corot-3b become constant toward their centers as has been previously found for Jupiter \citep{French2012}. However, Gliese-229b behaves differently and is characterized by 
a steady increase of the kinematic shear viscosity towards its center corresponding to the regions with densities above 100~g/cm$^3$.

\subsection{Electrical and Thermal Conductivity}\label{subsec:El_Theo}

The electrical $\sigma$ and thermal $\lambda$ conductivity are dominated by their electronic contributions, which
can be obtained within the framework of LRT using the frequency-dependent Onsager coefficients~\citep{Holst2011}
\begin{eqnarray}
 L_{mn}(\omega) = \frac{2\pi q^{4-m-n}}{3V m_e^2\omega}\sum_{\mathbf{k}\nu\mu}  |\langle\mathbf{k}\nu|\hat{\mathbf{p}}| \mathbf{k}\mu \rangle|^2
       (f_{\mathbf{k}\nu} - f_{\mathbf{k}\mu}) \nonumber\\
\times \!\left( \frac{E_{\mathbf{k}\mu}+E_{\mathbf{k}\nu}}{2}-h_e \!\right)^{m+n-2} \!\!\!\!\delta\!\left( E_{\mathbf{k}\mu} - E_{\mathbf{k}\nu} - \hbar\omega \right) .\;
\end{eqnarray}
The above equation contains the frequency $\omega$, the mass $m_e$ and charge $q=-e$ of the electron, the enthalpy per electron $h_e$, the eigenvalues $E_{\mathbf{k}\mu}$
with the Fermi-occupation number $f_{\mathbf{k}\mu}$ of the Bloch-state  $|\mathbf{k}\mu \rangle$ as well as the matrix elements
$\langle\mathbf{k}\nu|\hat{\mathbf{p}}| \mathbf{k}\mu \rangle$ with the momentum operator. According to this equation, there is a contribution to $L_{mn}(\omega)$ if the
energy of an incident photon $\hbar\omega$ is equal to the difference of two eigenstates. The Onsager coefficient $L_{11}(\omega)=\sigma(\omega)$ is given by the
frequency-dependent Kubo-Greenwood formula~\citep{Kubo1957,Greenwood1958}. The static limits of the electrical and thermal conductivity are obtained from their behavior
at $\omega\to0$:
\begin{eqnarray}
 \sigma  &=& \lim_{\omega\to 0} L_{11}(\omega)\label{eq:LimSig}\quad,\\
 \lambda &=& \lim_{\omega\to 0} \frac{1}{T}\left(L_{22}(\omega)-\frac{L_{12}^2(\omega)}{L_{11}(\omega)}\right)\quad. \label{eq:LimLam}
\end{eqnarray}

The ionic contributions to both quantities are neglected, since the considered objects are almost entirely composed of degenerate matter. In this regime both conductivities are governed
by the faster electrons and, hence, can be derived from static DFT calculations. For each density--temperature condition we averaged over 20 snapshots using a 4x4x4 Monkhorst-Pack
set to sample the Brillouin zone. Moreover, we used the PBE exchange-correlation functional~\citep{Perdew1996} throughout the calculation for the massive exoplanet and the two brown dwarfs, while the Jupiter
results~\citep{French2012} we compare to were obtained using the HSE exchange-correlation functional~\citep{Heyd2006}. Typically, it is necessary to use the more sophisticated HSE
exchange-correlation functional to describe the dissociation of hydrogen molecules appropriately. However, our three considered objects are dominated by degenerate matter with almost
no hydrogen bonds so that the PBE functional yields reasonable results.
Please note, that the addition of metals ($Z$) to the hydrogen-helium mixture would lead to slightly decreased conductivities due to stronger electron-ion scattering.

The results for the electrical and thermal conductivities are shown in the upper panels in Figure~\ref{fig:sigma} and Figure~\ref{fig:lambda},
respectively. Additionally, the associated quantities magnetic diffusivity,
\begin{equation}\label{eq:beta}
   \beta = \frac{1}{\sigma \mu_0} \quad,
\end{equation}
and thermal diffusivity,
\begin{equation}\label{eq:kappa}
   \kappa = \frac{\lambda}{\varrho c_p} \quad,
\end{equation}
can be calculated from the conductivities, where $\mu_0$ is the vacuum permittivity. Both quantities are included in Table~\ref{tab:Trans} and the thermal diffusivity is shown 
in the lower panel of Figure~\ref{fig:lambda}.

The electrical conductivity results are shown in Figure~\ref{fig:sigma}. Our DFT-MD results for KOI-889b (orange), Corot-3b (brown), and Gliese-229b (green) are
shown along with the model by \citet{Stevenson1977a} (dashed colored curves) as well as DFT-MD calculations for Jupiter (black line for HSE and open 
circles for PBE)~\citep{French2012}. The Jupiter results have been complemented by a prediction for the planet's rocky core shown as a dashed black curve. We assume that the core 
is made entirely of MgO and use a temperature of 20000~K and a density of 15~g/cm$^3$ as reference conditions~\citep{Cebulla2014}. Using the same approach as described
for the hydrogen-helium mixture, we obtain an electrical conductivity of roughly 90000 S/m. This corresponds to a bad metal {and thus supports the boundary condition to a
conducting core often made in magnetohydrodynamic simulations for Jupiter's magnetic field.

\begin{figure}[htb]
  \includegraphics[width=1.0\columnwidth]{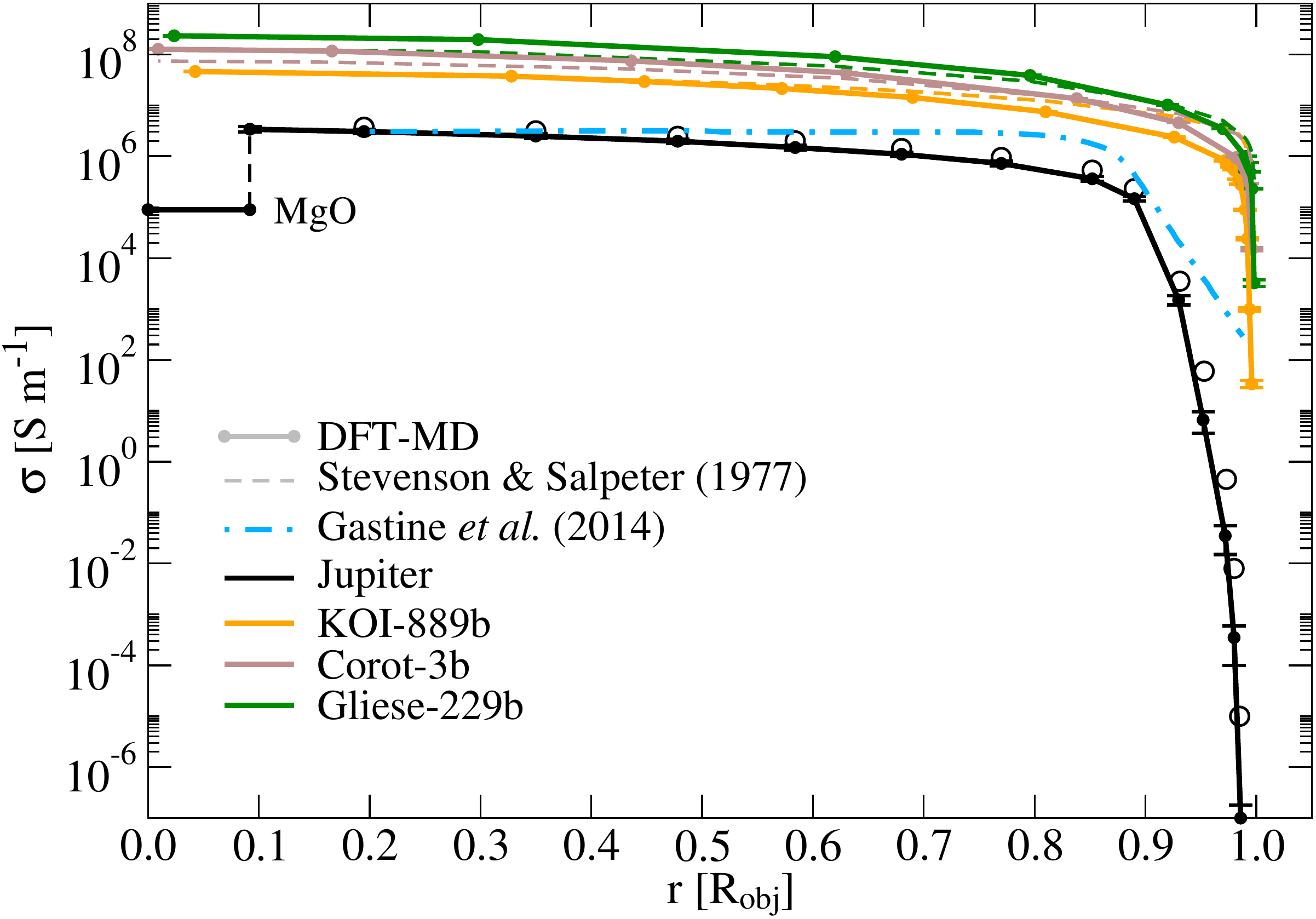}
  \caption{Electrical conductivity $\sigma$ along the considered isentropes. The results along the Jovian isentrope 
  derived from DFT-MD calculations using the HSE (black line) and PBE (open black circles) functional are taken from~\cite{French2012}. Furthermore, data by 
  \cite{Stevenson1977a} are shown.}
  \label{fig:sigma}
\end{figure}

The electrical conductivities from DFT-MD calculations have the steepest, super-exponential slope within the outermost 10\%, spanning a wide range of orders of magnitude. 
The values of our considered objects exceed the Jupiter results by at least 1.5 orders of magnitude. This is due to the significantly higher mass densities as well as
electron densities $n_e$ in the massive objects compared to Jupiter. The electrical conductivity in degenerate matter is proportional to the Fermi energy, which in turn
fulfills $E_F \sim n_e^{2/3}$. Additionally, helium is fully ionized under the conditions investigated in contrast to those present in Jupiter.

Comparing our results to \cite{Stevenson1977a}, we find especially the orange curve of KOI-889b to agree remarkably well with their predictions below \mbox{0.7 $R_{Obj}$.}
The electrical conductivity varies most in the interior of KOI-889b spanning four orders of magnitude. At the object's surface, as it is defined in section~\ref{sec:IntBrownie}, 
the hydrogen-helium mixture is not an insulator as has been predicted for Jupiter~\citep{French2012}. However, the conductivity is also not yet in the metallic range, so that this 
is most likely the region we introduce the largest error using the PBE instead of the HSE functional. This difference in functional amounts up to two orders of magnitude in the 
comparable region for Jupiter. Nevertheless, the electrical conductivity in KOI-889b as obtained with the PBE functional is eight orders of magnitude higher than the corresponding 
value in Jupiter at the same scaled radius. Furthermore, as shown by the Jupiter curves (see also \cite{French2012}, Figure~5), both exchange-correlation functionals agree 
well for the fully ionized regime. As the extent of the non-fully ionized region is negligibly small in the massive giant planet and the brown dwarfs, the utilization 
of the PBE instead of the HSE functional is justified.

Furthermore, Figure~\ref{fig:sigma} includes the conductivity data used in \cite{Gastine2014}. This combines an exponential branch for the strong decay in the outer shell 
with a polynomial branch for the interior. Note that numerical dynamo simulations use dimensionless formulations. For easy comparison we scaled the electrical conductivity 
of \cite{Gastine2014} by the corresponding value from \cite{French2012} for $R=0.2~\RJ$. Though the conductivity in the Jupiter dynamo models does not decay as steeply as 
predicted by \cite{French2012}, the Jovian model is nevertheless reproduced convincingly \citep{Gastine2014,Duarte2018}. The extreme gradient, that is better captured by 
the hyperbolic fit used in the dynamo models by \cite{Jones2014} and \cite{Cao2017}, has little additional effect on the dynamo process.

\begin{figure}[htb]
  \includegraphics[width=1.0\columnwidth]{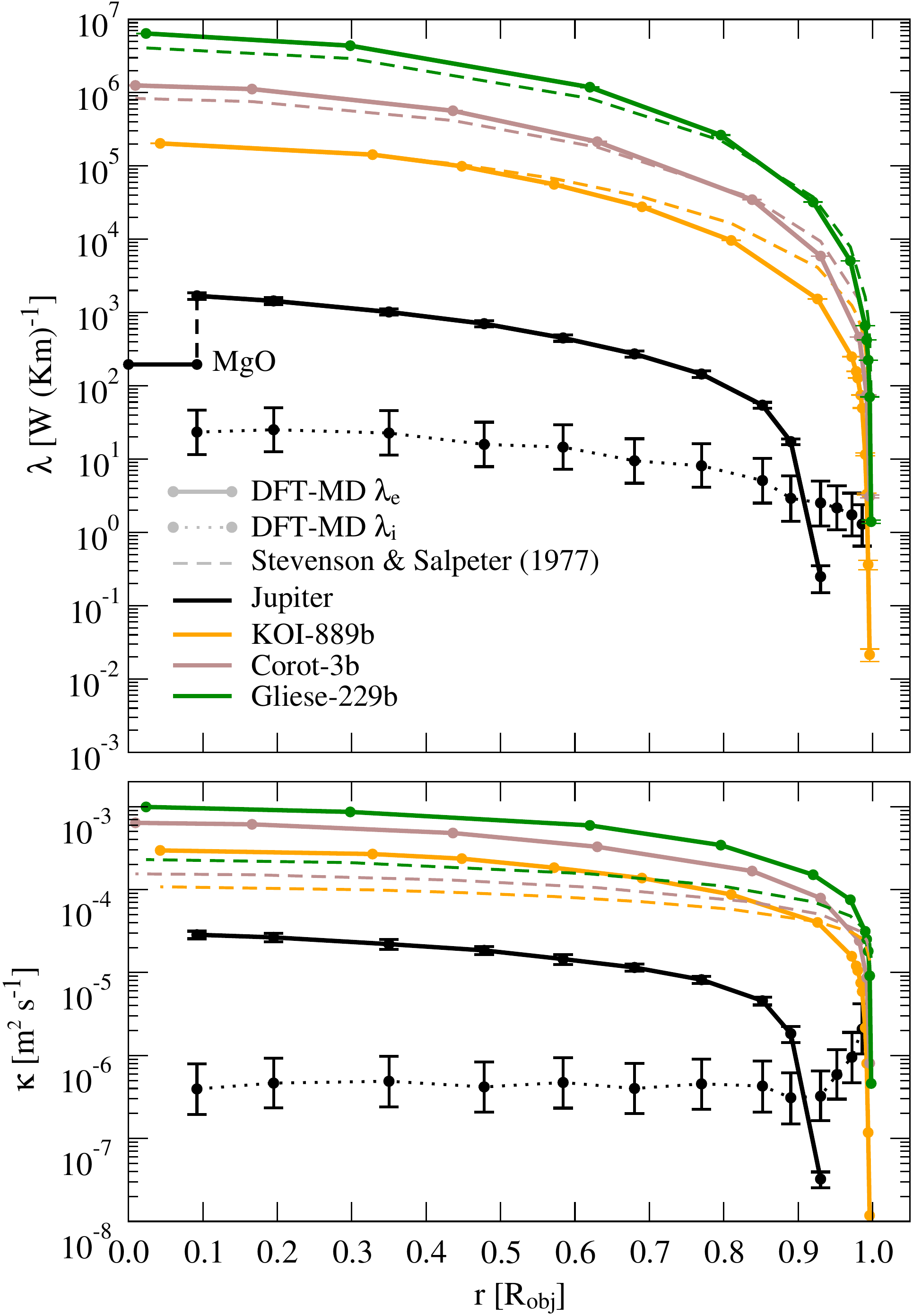}
  \caption{Thermal conductivity $\lambda$ (upper panel) and thermal diffusivity $\kappa$ (lower panel) along the considered isentropes
  (solid lines) in comparison to the results given by~\cite{Stevenson1977a} (dashed lines). The data for Jupiter, including the ionic 
  thermal conductivity $\lambda_i$ (dotted black line), were calculated by~\cite{French2012}.}
  \label{fig:lambda}
\end{figure}

The thermal conductivity is illustrated in the upper panel of Figure~\ref{fig:lambda} and resembles the general trend observed for the electrical conductivity. 
We find the thermal conductivity to exceed the respective values for Jupiter in the three considered objects at any given object radius. The slope of the colored 
curves is slightly steeper towards the objects' surfaces compared to Jupiter. Overall, the thermal conductivity is found to reach its smallest value at the surface
of KOI-889b. Moreover, our results are in good agreement with the earlier predictions by \citet{Stevenson1977a} (dashed curves). 
Hence, their approach  should work best for metals and degenerate matter as reflected by finding the smallest deviations between
both predictions for the most massive object Gliese-229b.

The dotted line in the upper panel of Figure~\ref{fig:lambda} shows the ionic contribution of the thermal conductivity in Jupiter~\citep{French2012} in contrast to
the solid lines that only contain the electronic contribution. Evidently, the ionic contributions dominate the non-metallic region near the surface of Jupiter. Such 
calculations, requiring the computation of heat flows~\citep{French2012}, have not been performed for the more massive objects, since their non-metallic region is very small.
Hence, we suggest to use the values provided for Jupiter by \cite{French2012} for the first outermost points of the massive objects. This is justified due to the 
similar thermodynamic conditions on the surface of all four objects, while the ionic contribution of the thermal conductivity varies only little with density and 
temperature. The same applies for the thermal diffusivity in the lower panel of Figure~\ref{fig:lambda}, which is the thermal conductivity scaled with the inverse of 
density and the specific heat capacity at constant pressure, see Equation~\eqref{eq:kappa}. The latter results show a significant deviation between our results and those
by \cite{Stevenson1977a}, who used a rather simple approximation for the thermal diffusivity. Their analytic approach assumes a constant heat capacity $c_p$ and only requires the
density as input.

\begin{deluxetable*}{ccccccccccc}
\tablecolumns{11}
\tablewidth{0pc}
\tablecaption{Linear transport properties of the massive giant planet KOI-889b as well as the brown dwarfs Corot-3b and Gliese-229b \label{tab:Trans}}
\tablehead{
\colhead{object} & \colhead{r} & \colhead{m} & \colhead{T} & \colhead{$\sigma$} & \colhead{$\beta$} 
& \colhead{$\lambda$} & \colhead{$\kappa$} & \colhead{$\eta$} & \colhead{$\nu$} & \colhead{$\kappa_R$}\\
\colhead{} & \colhead{[R$_{\mathrm{obj}}$]} & \colhead{[M$_{\mathrm{obj}}$]} & \colhead{[K]} & \colhead{[S m$^{-1}$]} & \colhead{[m$^2$ s$^{-1}$]} 
& \colhead{[W (Km)$^{-1}$]} & \colhead{[m$^2$ s$^{-1}$]} & \colhead{[mPas]} & \colhead{[mm$^2$ s$^{-1}$]} & \colhead{[cm$^2$ g$^{-1}$]}}
\startdata
KOI-889b & 0.996		& 0.99996		& 4800		& 33.7			& 23600		& 0.022		& 1.32$\times10^{-8}$		& 0.078		& 0.776		& 1.02$\times10^{4}$\\
\vdots & \vdots &\vdots &\vdots &\vdots &\vdots &\vdots &\vdots &\vdots &\vdots & \vdots\\
KOI-889b & 0.043		& 3.15$\times10^{-4}$	& 166000	& 4.61$\times10^{7}$	& 0.017		& 203020	& 2.91$\times10^{-4}$		& 23.56		& 0.506		& 3.10$\times10^{5}$\\
Corot-3b & 0.996		&	0.9558		& 7700		& 15000			& 53.05			& 3.100			& 7.66$\times10^{-7}$	& 0.149		& 0.719		& 5.61$\times10^{5}$\\
\vdots & \vdots &\vdots &\vdots &\vdots &\vdots &\vdots &\vdots &\vdots &\vdots & \vdots\\
Corot-3b & 9.29$\times10^{-3}$ 	& 3.07$\times10^{-5}$	& 420000	& 1.265$\times10^{8}$	& 6.29$\times10^{-3}$	& 1.25$\times10^{6}$	& 6.27$\times10^{-4}$	& 92.78		& 0.679		& 4.38$\times10^{4}$\\
Gliese-229b & 0.998		&	1.0000			& 8000			& 3200			& 245.6			& 1.40			& 4.60$\times10^{-7}$	& 0.136		& 1.080		& 3.88$\times10^{5}$\\
\vdots & \vdots &\vdots &\vdots &\vdots &\vdots &\vdots &\vdots &\vdots &\vdots & \vdots\\
Gliese-229b & 0.024 		&	6.84$\times10^{-5}$	& 1.18$\times10^{6}$	& 2.32$\times10^{8}$	& 3.42$\times10^{-3}$	& 6.42$\times10^{6}$	& 8.83$\times10^{-4}$	& 581.0		& 1.300		& 3000\\
\enddata
\tablenotetext{}{(This table is available in its entirety in machine-readable form.)}
\end{deluxetable*}

\subsection{Lorenz number}

To check whether our results obey the theoretical predictions for a degenerate electron gas, we calculate the Lorenz
number~\citep{Lorenz1872}, which is defined by the Wiedemann-Franz law
\begin{equation}
  L=\frac{e^2}{k_B^2T}\frac{\lambda}{\sigma} \quad.
\end{equation}

\begin{figure}[htb]
  \includegraphics[width=1.0\columnwidth]{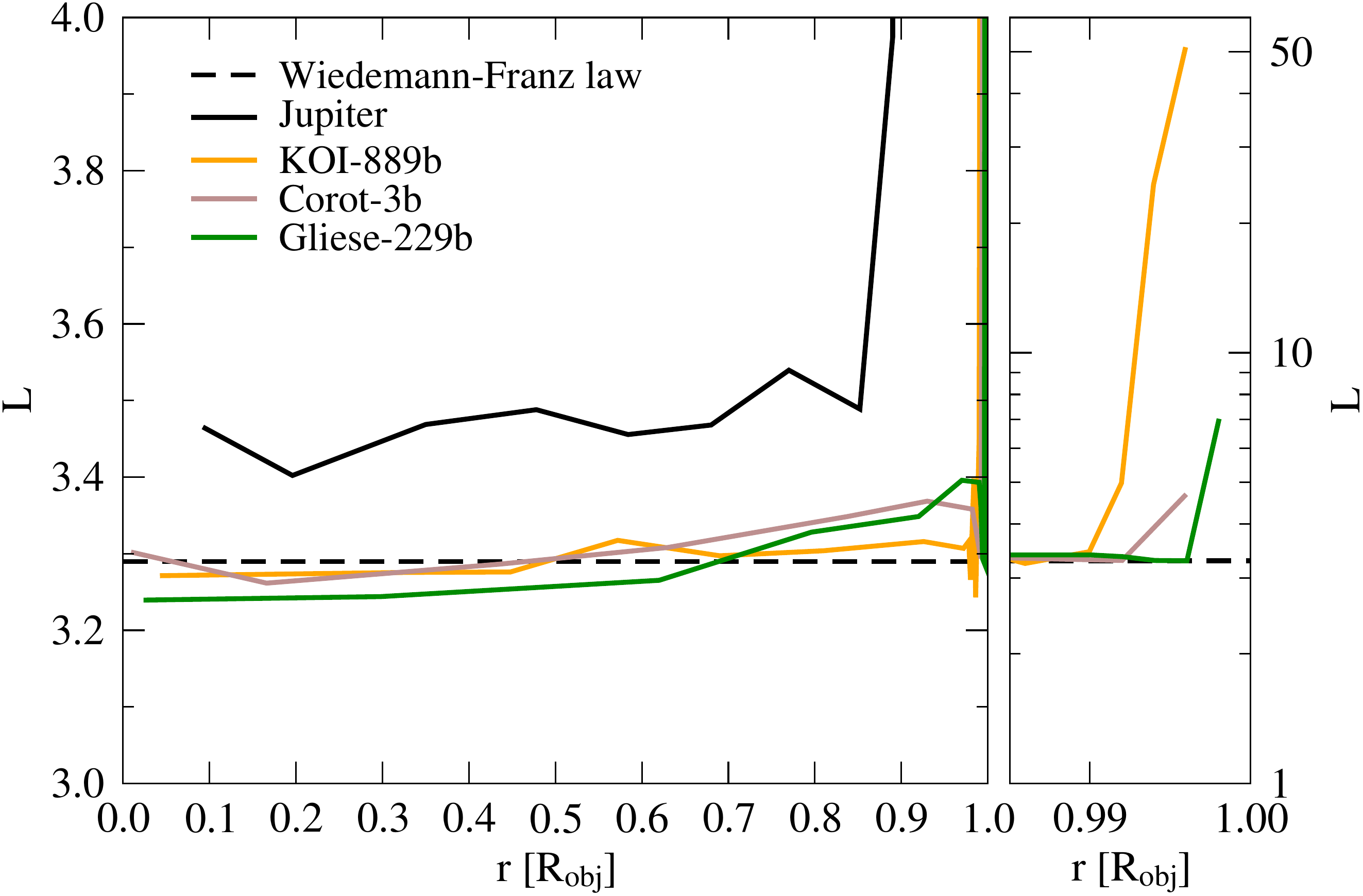}
  \caption{The left panel shows the Lorenz number along the isentropes (solid lines) of Jupiter (black)~\citep{French2012}, KOI-889b (orange), Corot-3b (brown), and 
  Gliese-229b (green). The Wiedemann-Franz limit (dashed black line) is shown as well. 
  The right panel shows the Lorenz number near the surface of the considered objects.}
  \label{fig:lorenz}
\end{figure}

The results are shown in Figure~\ref{fig:lorenz}. The Wiedemann-Franz law of $L=\pi^2/3$ is strictly valid only for the fully degenerate limit
$\Theta\ll 1$. The results for Jupiter are about 6~\% higher, most likely due to the contribution of neutral helium atoms which are not ionized under these conditions.
The exponential increase near the surfaces can be addressed to the transition from a liquid metal to a partially ionized plasma along the isentrope. 
For more compact objects this process is shifted towards the outer regions.
Such a behavior of the Lorenz number for warm dense hydrogen has already been shown 
by \cite{Holst2011} for decreasing densities and temperatures (see \fig 8 therein).

\subsection{Opacity}\label{subsec:Opaz_Result}

Further optical properties can be determined from the frequency-dependent electrical conductivity~\citep{French2011}. The imaginary part of the complex electrical
conductivity $\sigma(\omega)=\sigma_1(\omega)+ \mathrm{i}\sigma_2(\omega)$ can be calculated using a Kramers-Kronig relation. The complex dielectric function
$\epsilon(\omega)=\epsilon_1(\omega)+ \mathrm{i}\epsilon_2(\omega)$ then emerges from $\epsilon_1(\omega) = 1-\sigma_2(\omega)/\epsilon_0\omega$ and
$\epsilon_2(\omega) = 1-\sigma_1(\omega)/\epsilon_0\omega$. For absorption features one needs to determine the complex refraction index via
$\epsilon(\omega)$: $n(\omega) + \mathrm{i}k(\omega) = \sqrt{\epsilon({\omega})}$.
The extinction-coefficient $k(\omega)$ and the speed of light $c$ yield the frequency-dependent absorption coefficient $\alpha(\omega)=2\omega k(\omega)/c$. Finally, we
obtain the Rosseland mean opacity $\kappa_R$ as a measure for the optical thickness of a system~\citep{Kippenhahn1991},
\begin{equation}
 \frac{1}{\kappa_R} = \frac{15}{4\pi^4} \int \limits_0^{\infty} \frac{d\omega}{\alpha(\omega)}\frac{x^4\mathrm{e}^x}{(\mathrm{e}^x-1)^2} \quad,
\end{equation}
with the substitution $x=\hbar\omega/(k_BT)$.
The inverse absorption coefficient is weighted with the derivative of the Planck function with respect to temperature. The resulting expression is integrated over all frequencies.
$\kappa_R$ is an important quantity to determine, e.g., for describing radiation transport inside the objects. If the matter is optically thick, energy cannot be transported 
efficiently via radiation and convection becomes the dominant transport mechanism (Schwarzschild criterion). The following results for $\kappa_R$ from DFT-MD simulations only contain the 
electronic contributions. Rotational and vibrational excitations are neglected, which is justified for most of the interior (fully dissociated and ionized). However, the strong absorption 
of molecules, especially in the cool atmospheres of giant planets and brown dwarfs, is not considered. Likewise, we neglect absorption from heavier elements 
(oxygen, carbon, iron and noble gases) since the simulations are based on a representative mixture of hydrogen and helium.

\begin{figure}[htb]
  \includegraphics[width=1.0\columnwidth]{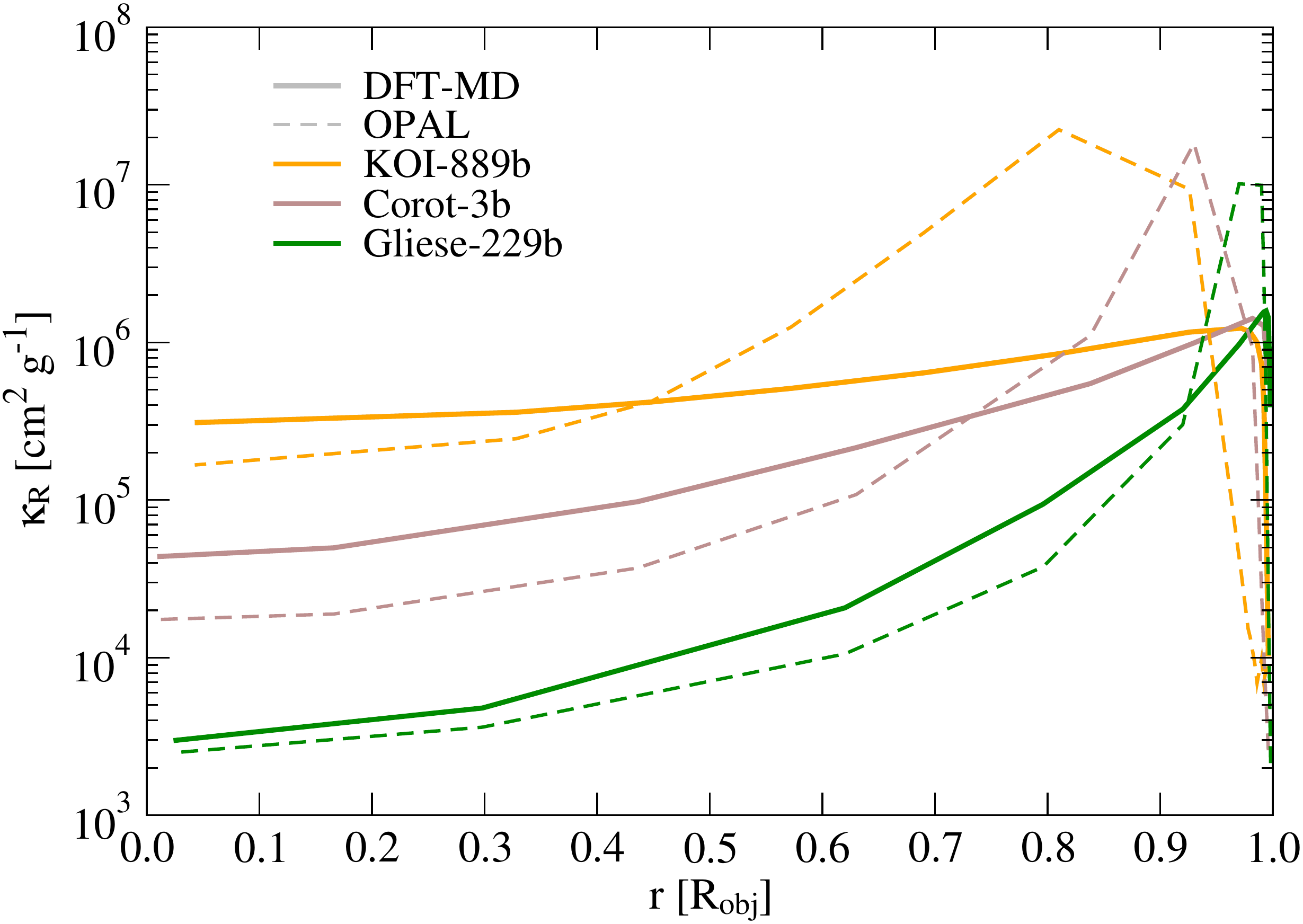}
  \caption{Rosseland mean opacity $\kappa_R$ derived from DFT-MD calculations (solid curves) for KOI-889b (orange), Corot-3b (brown), and Gliese-229b (green). 
  The respective values from the OPAL tables (dashed curves:~\cite{Rogers1998}) are shown as well.}
  \label{fig:opaz}
\end{figure}

In Figure~\ref{fig:opaz}, we compare our results (solid curves) to those from the OPAL tables~\citep{Rogers1994,Rogers1998} (dashed curves) obtained from interpolation of those
tables with respect to the thermodynamic conditions along the isentropes of the objects. We find the highest opacities if the system contains bound electrons and the main contribution
arises from bound-free transitions. Similar to the electrical conductivity this is the case in the outermost parts of KOI-889b and the brown dwarfs, where the conductivities 
increase in a super-exponential way. The thermodynamic states in the interior of an object become more extreme and thus more ionized with increasing mass. 
Hence, the maximum of $\kappa_R$ is closer to the surface than for less massive objects due to bound-free transitions. Therefore Gliese-229b (green curves) has the outermost location 
of the maximum of $\kappa_R$ compared to Corot-3b (brown curves) and KOI-889b (orange curves).
The higher the pressure in the interior, the faster helium ionizes.
Consequently, after the ionization of hydrogen bound-free transitions in helium disappear faster for massive objects, leading to a stronger slope of the opacity. Therefore, the slope 
of $\kappa_R$ inside KOI-889b towards the center is smaller than for Corot-3b and Gliese-229b. Qualitatively, the OPAL results show the same behavior as the DFT-MD curves but with a 
more pronounced maximum. 
OPAL treats the various absorption processes based on atomic physics
codes and takes into account plasma effects such as line broadening and
screening in an approximate way. The DFT-MD method considers these
effects within the electronic structure calculations for given ion
configurations consistently.

\section{Implications for planetary and stellar dynamos}\label{sec:Implications}

The Jupiter model data by \cite{French2012}, which served as a reference in the discussion above, have been used in numerical simulations of Jupiter's dynamo successfully reproducing
the planet's large scale magnetic field \citep{Gastine2014,Jones2014,Duarte2018}. A particularly interesting feature is the dynamo action of the zonal winds that are driven by Reynolds 
stresses acting in the outer molecular layer. Where these winds reach down to sizable electrical conductivities, they modify the large scale field produced by a primary dynamo at greater 
depth \citep{Gastine2014}. This secondary dynamo creates banded structures at low latitude that have recently been confirmed by the Juno mission \citep{Connerney2018}. The numerical 
simulations suggest that these bands are the expression of pole-ward propagating waves \citep{Gastine2012}.

The properties derived here suggest that the dynamics in large exoplanets or brown dwarfs may be very similar to the scenario explored for Jupiter. The molecular hydrogen envelope 
occupies a smaller fraction and has therefore largely not been considered here. We expect that, just like on Jupiter, these objects will harbor fierce zonal wind systems and the 
associated dynamo action. Concerning the deeper dynamics, it seems remarkable that the kinematic viscosity remains not only surprisingly similar to that in Jupiter but also remains 
more or less homogeneous and is thus virtually independent of the density. Moreover, since the thermal diffusivity increases relatively mildly with density while the opacity is huge, 
convective motions will likely remain the preferred mode of heat transport. Since exoplanets or brown dwarfs also seem to be relatively fast rotators with decent luminosities, their 
internal dynamos may operate in the same regime as, for example, Jupiter or many fully convective stars \citep{Reiners2010}.

\cite{Christensen2009} and \cite{Yadav2013} show that the field strength of such objects is successfully predicted by a scaling law that formulates a simple dependence 
on the luminosity. For a typical brown dwarf, this law would predict a field of about $0.2\,$T \citep{Reiners2010}. Radio emissions confirm that several brown dwarfs 
indeed possess  a magnetic field but indicate that the field strengths are somewhat larger \citep{Kao2016}. Recently, the first detection of Zeeman line broadening for 
a brown dwarf suggests a field of about $0.5\,$T, covering at least $10\,$\% of the surface.

The reason why the scaling laws seem to underpredict the field strength remains unclear and may simply be an expression of the uncertainties  \citep{Kao2016}. 
However, a dynamo simulation by \citet{Yadav2015} offers an alternative explanation. Dynamo simulations with a small Prandtl number, the ratio of kinematic viscosity to 
thermal diffusivity, show particularly strong localized surface field patches on top of a weaker larger scale background field. Such a combination has been observed 
for several low mass stars \citep{Reiners&Basri2009}. Zeeman line broadening as well as radio emission data may predominantly constrain the strong localized 
surface field while scaling laws mostly concern the deeper produced global field.

In Jupiter, the magnetic Prandtl number becomes as small as $10^{-2}$ at depth \citep{French2012}. The data presented above suggest even smaller values in more massive objects, 
for example values down to $1.27\cdot10^{-3}$ in Gliese-229b. Dedicated dynamo simulations based on the properties presented here will be required to explore the possible 
particularities of the dynamics and magnetic fields in massive exoplanets and brown dwarfs.

\section{Conclusions}\label{sec:Conclusion}

We have determined the thermophysical properties of H-He mixtures for conditions inside massive giant planets and brown dwarfs based on \textit{ab initio}
simulations. In particular, we discussed thermodynamic material properties, the Love number, the equidistance, as well as transport
properties including the closely related opacity. The provided values represent a considerable extension of the dataset calculated for Jupiter by \cite{French2012}.
In comparison to Jupiter, the underlying models \citep{Becker2014} start at greater pressures and temperatures where most of the hydrogen is already dissociated. The 
thermodynamic material properties and transport properties 
therefore largely lack the features that characterize the properties in Jupiter's outer envelope. For example, the extreme rise in electrical conductivity and the 
dissociation maximum in the heat capacities \citep{French2012} are absent for the considered massive objects.
The properties of degenerate matter play an increasingly important role when the object mass grows.

Overall, our dataset, combined with the Jupiter data by \cite{French2012}, increases our knowledge of extreme thermodynamic conditions, covering the
broad mass range from Jupiter-sized giant planets up to brown dwarfs.
This data will stimulate the development of 
new models for the interior structure, thermal evolution and internal dynamics of massive exoplanets and brown dwarfs.

\acknowledgments
We thank N.\ Nettelmann, M.\ French, D.\ Cebulla, M.\ Sch\"ottler, W.\ Lorenzen, and S.\ Hamel for helpful discussions and the anonymous referee for providing useful comments.
This work was supported by the Deutsche Forschungsgemeinschaft within the SFB~652 and FOR 2440.
The DFT-MD simulations were performed at the North-German Supercomputing Alliance (HLRN)
and at the IT and Media Center of the University of Rostock. Mandy Bethkenhagen further acknowledges support by the U.S.\ Department
of Energy at the Lawrence Livermore National Laboratory under Contract No.\ DE-AC52-07NA27344.

\bibliographystyle{apj}
\bibliography{statphys.bib}

\end{document}